\documentclass{PoS}

\newcommand{\be}{\begin{equation}}
\newcommand{\ee}{\end{equation}}
\newcommand{\bary}{\begin{eqnarray}}
\newcommand{\eary}{\end{eqnarray}}

\title{The early afterglow and magnetized ejecta present in  GRB 110731A}

\ShortTitle{GRB110731A}

\author{\speaker{Nissim Fraija}\thanks{Luc Binette postdoctoral scholarship.}\\
        Instituto de Astronomía, UNAM\\
        E-mail: \email{nifraija@astro.unam.mx}}

\author{William H. Lee\\
        Instituto de Astronomía, UNAM\\
        E-mail: \email{wlee@astro.unam.mx}}

\abstract{One of the most energetic gamma-ray bursts GRB 110731A, was observed from optical to GeV energy range by Fermi and Swift Observatories, and by the MOA and GROND optical telescopes.  The multiwavelength observations over different epochs (from trigger time to more than 800 s) showed that the spectral energy distribution was better fitted by a wind afterglow model.     We present a leptonic model  based on an early afterglow that evolves in a stellar wind to describe the multiwavelength light curves observations. In particular, the origin of the LAT emission is explained through the superposition of synchrotron radiation from the forward shock and  synchrotron self-Compton emission from the reverse shock. The bulk Lorentz factor required in this model is  $\Gamma\simeq520$ and the result suggests that the ejecta must be magnetized.}

\FullConference{Swift: 10 Years of Discovery,\\
		2-5 December 2014\\
		La Sapienza University, Rome, Italy }

\begin{document}

\section{Introduction}
In recent years,  the detection of $\gamma$-ray and optical  polarization in gamma-ray bursts  (GRBs)  has supported the idea that jets could be magnetized \cite{2003Natur.423..415C}.  The jet evolution with  magnetic  content has been explored in several contexts. In these models, an electromagnetic component  is introduced through the magnetization parameter ($\sigma$) and  defined by the ratio of Poynting flux (electromagnetic component) and matter energy (internal+kinetic component) \cite{2002A&A...387..714D,2003ApJ...596.1104V}.\\ 
The afterglow phase is one of the most interesting and least understood episodes of the burst. During this episode, the relativistic ejecta interacts with the surrounding matter generating reverse and forward shocks.   In a wind and homogeneous environment, the reverse shock  has been discussed to describe the early $\gamma$-ray, optical and/or radio flares present in some bursts \cite{2003ApJ...597..455K, 2012ApJ...751...33F,2012ApJ...755..127S,2003ApJ...589L..69L} and  the forward shock to explain the continuous softening of the afterglow spectrum \cite{2007MNRAS.379..331P}.\\ 
Hadronic (inelastic proton-neutron collisions and proton-photon interactions)  and leptonic (inverse Compton (IC), synchrotron self-Compton (SSC) and synchrotron emission) models  have been widely explored to explained the photons observed with energies $\geq$ 100 MeV \cite{ 2004A&A...418L...5D,2001ApJ...548..787S, 2001ApJ...559..110Z}.\\ 
The bright and long  GRB 110731A was observed in $\gamma$-ray, X-ray and optical wavelengths.  The analysis of the prompt phase  revealed an extremely bright peak and a temporally extended component  in the  Large Area Telescope (LAT)  light curve  starting at $\sim$ 5.5 s \cite{2013ApJ...763...71A}.  In addition, temporal and spectral analysis in different wavelengths and epochs  (just after the trigger time and extending for more than 800 s) favored a wind afterglow  model.   Recently, Lemoine et al.  (2013) \cite{2013arXiv1305.3689L} proposed that if the magnetic equipartition parameter would vary as a function of time, $\epsilon_B\propto t^{-\alpha_t}$ with  $0.5\leq\alpha_t \leq 0.4$, then the GeV photons would be  likely created in a region of strong $\epsilon_B$. They argued that the magnetization that permeates the blast wave of GRB 110731A could be described as partial decay of the micro-turbulence \cite{2003MNRAS.339..881R} as observed in particle-in-cell (PIC) simulations \cite{2011ApJ...726...75S}.\\
In this paper, we develop a leptonic model  based on external shocks (forward and reverse) that evolves adiabatically in a stellar wind.  We show that with the suitable equipartition parameters we are able to reproduce the LAT, X-ray and optical light curves (LCs) observed in GRB 110731A.
\section{GRB 110731A}
GRB 110731A was localized with coordinates R. A.=18$^h$41$^m$00$^s$ and dec.=-28$^\circ$31'00'' (J2000), with a 68\% confidence error radius of 0.2$^\circ$.

This burst was detected by the Gamma-Ray Burst Monitor and LAT \cite{2013ApJ...763...71A} on board Fermi; BAT, XRT and UVOT \cite{2011GCNR..343....1O}  on board Swift, and the Microlensing Observations in Astrophysics (MAO) telescope  \cite{2011GCN..12225...1T} and Gamma- ray Burst Optical/Near-Infrared Detector (GROND).  LAT began to observe this burst from $\sim$ 4s after trigger time to more than 800 s, detecting an extremely bright peak at 5.5 s. Swift/BAT perceived immediately this burst after the detection  by both instruments of Fermi, whereas XRT and UVOT began observations 56s after the BAT trigger.  UVOT swiftly determined the afterglow position as R.A. = 18$^h$42$^m$00$^s$.99 and dec.=-28$^\circ$32'13''.8 (J2000), with a 90\% confidence. The lack of observation in the UV filters is consistent with the measured redshift z=2.83 \cite{2011GCN..12225...1T}. MAO observations started 3.3 minutes after the Swift trigger and  GROND  mounted on the 2.2 m MPG/ESO telescope at La Silla Observatory, Chile, observed this burst 2.74 days after the trigger  \cite{2008PASP..120..405G}.\\

\section{Wind-afterglow Model}
Afterglow hydrodynamics involves a relativistic blast wave expanding into the medium with density
\be\label{rho}
\rho=A\,r^{-2} \hspace{0.3cm} {\rm with}  \hspace{0.3cm}  A=\frac{\dot{M}_w}{4\pi V_w}\,,
\ee
where $\dot{M}_w$ is the mass loss rate and $V_w$ is the wind velocity.  We hereafter use primes (unprimes) to define the quantities in a comoving (observer) frame and  c=$\hbar$=1 in natural units. The radius shock ($r$) spreading into this density can be written in the form
\be\label{rad}
r=\frac{3\xi}{2\pi^{1/2}} (1+z)^{-1/2}\,E^{1/2}\,t^{1/2}\,A^{-1/2}\,.
\ee
where the total energy of the shock is $E=8\pi/9\,A\,\Gamma^2\,r$ with $\Gamma$ the bulk Lorentz factor, $\xi$ a constant parameter \cite{1998ApJ...493L..31P}, $z$ the redshift and $t=(1+z)\,\frac{r}{4\,\xi^2\,\Gamma^2}$  is the time in the observer's frame.\\

By considering the typical values of the stellar wind ($A=A_{\star}\times\,(5.0\times 10^{11})$ g/cm with $A_{\star}=0.1$; \cite{2000ApJ...536..195C}, the parameter $\xi=0.56$; \cite{1998ApJ...493L..31P}) and those inferred by observations: redshift  $z=2.83$ \cite{2011GCN..12225...1T}, total energy $E\simeq10^{54}$ erg and duration of GRB $T_{90}$= 7.3 s, we calculate the SSC and synchrotron spectral  breaks from forward and reverse shocks. The subscripts f and r refer throughout this paper to the forward and reverse shocks, respectively.
\subsection{Forward Shocks}
By considering the brightest peak of the flux density present at the end of the prompt emission, in the interval [5.47 s, 5.67 s], and also the LAT flux decaying smoothly during the whole temporally extended emission, we constrain the Lorentz factor $\Gamma\simeq 520$ so that the deceleration time occurs at
\be\label{tdec}
t_{dec}\simeq5.55\,{\rm s}\biggl(\frac{1+z}{4}\biggr)\,E_{54}\,A^{-1}_{\star,-1}\,\Gamma^{-4}_{2.72}\,.
\ee
We consider that electrons are accelerated in the shock to a power-law distribution  $N(\gamma_e) d\gamma_e\propto \gamma_e^{-p} d\gamma_e$ with a minimum Lorentz factor $\gamma_e\leq \gamma_m=\epsilon_{e,f}(p-2)/(p-1)\,m_p/m_e \Gamma$, where $\epsilon_B=B^2/(32\pi\Gamma^2 \rho)$ and $\epsilon_e=U_e/(4\Gamma^2\rho)$ are the magnetic and electron equipartition parameters, respectively, and the magnetic field is ${\small B'_f\simeq \frac{8\sqrt2\,\pi}{3\xi}(1+z)^{1/2} \epsilon_{B,f}^{1/2}\,\Gamma\,E^{-1/2}\,t^{-1/2}\,A}$.  Comparing the cooling  {\small $t_{e,syn}\simeq 3m_e/(16\sigma_T)\,(1+x_f)^{-1}\,(1+z)\,\epsilon^{-1}_{B,f}\,\rho^{-1}\,\Gamma^{-3}\,\gamma_e^{-1}$} and  acceleration $t_{acc}\simeq \frac{2\pi\,m_e}{q_e}(1+z)\,\Gamma^{-1}\,{B'}^{-1}_f \gamma_e$ time scales with the deceleration time (eq. \ref{tdec}), we obtain the  cooling ${\small \gamma_{e,c,f}=\frac{3m_e\xi^4}{\sigma_T}\,(1+x_f)^{-1}\,(1+z)^{-1}\,\epsilon^{-1}_{B,f}\,\Gamma\,A^{-1}\,t}$ and maximum ${\small \gamma_{e,max,f}\simeq \sqrt{\frac{9\sqrt2\,q_e}{16\,\pi \sigma_T}}\,\xi^{1/2}\,(1+z)^{-1/4}\epsilon_{B,f}^{-1/4}\Gamma^{-1/2}E^{1/4}
A^{-1/2}\,t^{1/4}}$  electron Lorentz factors.  With the previous relations, we can write the synchrotron spectral breaks as
\bary\label{synforw_a}
E^{syn}_{\rm \gamma,a,f} &\simeq&  5.56\times 10^{-4}\,{\rm eV}\, \biggl(\frac{1+z}{4}\biggr)^{-2/5}\,\epsilon_{e,f,-0.4}^{-1}\,\epsilon_{B,f,-4.15}^{1/5}\,A^{6/5}_{\star,-1}\,E^{-2/5}_{54}\,  t_{1}^{-3/5}  \cr
E^{syn}_{\rm \gamma,m,f} &\simeq&  77.45\,{\rm keV}\, \biggl(\frac{1+z}{4}\biggr)^{1/2}\,\epsilon_{e,f,-0.4}^2\,\epsilon_{B,f,-4.15}^{1/2}\,E^{1/2}_{54}\,  t_{1}^{-3/2}\cr
E^{syn}_{\rm \gamma,c,f}  &\simeq&  0.30\, {\rm eV}\, \biggl(\frac{1+z}{4}\biggr)^{-3/2}\,\biggl(\frac{1+x_f}{11}\biggr)^{-2}\, \epsilon_{B,f,-4.15}^{-3/2}\,A^{-2}_{\star,-1} E^{1/2}_{54}\, t_{1}^{1/2}\, \cr
E^{syn}_{\rm \gamma,max,f}  &\simeq&36.94\, {\rm GeV} \,\biggl(\frac{1+z}{4}\biggr)^{-3/4}\,E^{1/4}_{54}\,A^{-1/4}_{\star,-1}\,t^{-1/4}_1\,.
\eary
\noindent Here $\sigma_T$ is the Thomson cross section, $m_e$ is electron mass, $q_e$ is the elementary charge, the term $(1+x_f)$ is introduced as a correction factor and $D$ is the luminosity distance.  The transition time from fast- to slow-cooling regime is $t^{syn}_0=123.46 s \,\bigl(\frac{1+z}{4}\bigr)\epsilon_{e,f,-0.4}\,\epsilon_{B,f,-4.15}\,A_{\star,-1}$.  It is worth noting that X-ray and optical fluxes  can be described through the spectral evolution of the characteristic  $E^{syn}_{\rm \gamma,m,f}\simeq  77.45\,{\rm keV}\ $ and cooling $E^{syn}_{\rm \gamma,c,f}  \simeq  0.30\, {\rm eV}$ break energies and after $\sim 100$ s the synchrotron spectrum changes from fast to slow regime.  From the maximum photon energy $E^{syn}_{\rm \gamma,max,f}  \simeq 36.94\, {\rm GeV}$, one can see that  the temporally extended LAT component can be explained by synchrotron emission of  high-energy electrons radiating at $\simeq$ 100 MeV.  
%
%It is important to highlight that the maximum photon energy achieved by synchrotron radiation is $E^{syn}_{\rm \gamma,max,f} \simeq 36.94\, {\rm GeV}$ for t=10s and $E^{syn}_{\rm \gamma,max,f} \simeq 20.77\, {\rm GeV}$ for t=100s.\\
%
%\subsubsection{SSC emission}
%
Fermi-accelerated electrons may scatter synchrotron photons up to higher energies as $E^{ssc}_{\gamma,i}\simeq 2 \gamma^2_{e,i} E^{syn}_{\gamma,i}$. From the synchrotron spectral breaks (eqs. \ref{synforw_a}), the spectral breaks in the Compton regime are
\bary\label{sscforw_a}
E^{ssc}_{\rm \gamma, m,f} &\simeq&  11.66\,{\rm TeV}\biggl(\frac{1+z}{4}\biggr)\,\epsilon_{e,f,-0.4}^4\,\epsilon_{B,f,-4.15}^{1/2}\,A^{-1/2}_{\star,-1}\,E_{54}\,  t_{2}^{-2}\cr
E^{ssc}_{\rm \gamma, c,f}  &\simeq&  162.8\, {\rm keV}\, \biggl(\frac{1+z}{4}\biggr)^{-3}\,\biggl(\frac{1+x_f}{11}\biggr)^{-4}\,\epsilon_{B,f,-4.15}^{-7/2}\, A^{-9/2}_{\star,-1}\,E_{54}\, t_{2}^{2}\, \cr
F^{ssc}_{\rm \gamma, max,f}&\simeq&  0.41\,{\rm Jy}\biggl(\frac{1+z}{4}\biggr)\,\epsilon_{B,f,-4.15}^{1/2} \,A^{1/2}_{\star,-1}\,D^{-2}_{28}E_{54}\,,
\eary
where the break energy at the Klein-Nishina regime is $E^{KN}_{\gamma,f}= 42.33\, {\rm GeV}$.  The value of equipartition parameters $\epsilon_{B,f}=10^{-4.15}$ and $\epsilon_{e,f}=0.4$ were obtained after fitting the multiwavelength LC observations (see \cite{Fraija}).  We can see that the maximum photon energy achieved by synchrotron radiation is  36.94 GeV and the characteristic SSC spectral break is 11.66 TeV, therefore the photon with energy of 3.4 GeV detected at hundreds of seconds \cite{2013ApJ...763...71A} could be explained by the synchrotron and/or SSC emission. 

\subsection{Reverse shocks}
%A simple analytic solution can be derived taking two limiting cases, thick- and thin-shell case,  
The critical Lorentz factor can be written as 
\bary\label{gamma_cr}
\Gamma_c&\simeq&472.5\biggl(\frac{1+z}{4}\biggr)^{1/4} \,A^{-1/4}_{\star,-1}\,E^{1/4}_{54}\left(\frac{T_{90}}{7.3s}\right)^{-1/4}\,.
\eary
Since the brightest peak was present in the interval [5.47 s, 5.67 s], we consider that bulk Lorentz factors at forward and reverse shocks are equal $\Gamma\simeq 520 > \Gamma_c$, then the reverse shock evolves in the thick-shell case.
% \cite{2003ApJ...597..455K} 
%\subsubsection{Synchrotron emission}
% 
 Synchrotron spectral breaks between forward and reverse shocks are related by \cite{2003ApJ...597..455K} 
\bary\label{conec}
E^{syn}_{\rm \gamma, m,r}\sim\,\mathcal{R}^2_e\,\mathcal{R}^{-1/2}_B\,\mathcal{R}^{-2}_M\,E^{syn}_{\rm \gamma,m,f},\,\,\,\,E^{syn}_{\rm \gamma,c,r}\sim\,\mathcal{R}^{3/2}_B\,\mathcal{R}^{-2}_x\,E^{syn}_{\rm \gamma,c,f},\,\,\,\,\,\,F^{syn}_{\rm \gamma,max,r}\sim\,\mathcal{R}^{-1/2}_B\,\mathcal{R}_M\,F^{syn}_{\gamma,max,f}
\eary
where  ${\small  \mathcal{R}_B=\frac{\epsilon_{B,f}}{\epsilon_{B,r}},\hspace{0.1cm} \mathcal{R}_e=\frac{\epsilon_{e,r}}{\epsilon_{e,f}},\hspace{0.1cm}  \mathcal{R}_x=\frac{1+x_f}{1+x_r+x_r^2}\hspace{0.1cm} {\rm and}\hspace{0.1cm} \mathcal{R}_M=\frac{\Gamma^2_{c}}{\Gamma}}$.   From eqs. (\ref{synforw_a}, \ref{conec} and \ref{gamma_cr}), we get that the synchrotron break energies can be written as
\bary\label{synrev_a}
E^{syn}_{\rm \gamma,a,r}&\simeq& 4.28\times 10^{-8} \, {\rm eV}\, \biggl(\frac{1+z}{4}\biggr)^{-7/5}\,\epsilon_{e,r,-0.4}^{-1}\,\epsilon_{B,r,-0.55}^{1/5}\,\Gamma^{2}_{2.72}\, A_{\star,-1}^{11/5}\,E^{-7/5}_{54}\,\left(\frac{T_{90}}{7.3s}\right)^{2/5} \cr
E^{syn}_{\rm \gamma,m,r}&\simeq& 128.94  \, {\rm eV}\, \biggl(\frac{1+z}{4}\biggr)^{-1/2}\,\epsilon_{e,r,-0.4}^{2}\,\epsilon_{B,r,-0.55}^{1/2}\,\Gamma^{2}_{2.72}\, A_{\star,-1}\,E^{-1/2}_{54}\,\left(\frac{T_{90}}{7.3s}\right)^{-1/2} \cr
E^{syn}_{\rm \gamma,c,r}&\simeq& 0.93\times 10^{-5} \, {\rm eV}\,  \biggl(\frac{1+z}{4}\biggr)^{-3/2}\,\biggl(\frac{1+x_r+x_r^2}{3}\biggr)^{-2}\,\epsilon_{B,r,-0.55}^{-3/2}\,A^{-2}_{\star,-1}\,E^{1/2}_{54}\,\left(\frac{T_{90}}{7.3s}\right)^{1/2} \cr
F_{\rm \gamma,max,r}&\simeq&  4.31\times10^4  \,{\rm \,Jy}\,  \biggl(\frac{1+z}{4}\biggr)^{2}\,\epsilon_{B,r,-0.55}^{1/2}\,\Gamma^{-1}_{2.72}\,A^{1/2}_{\star,-1}\,D^{-2}_{28}\,E_{54}\,\left(\frac{T_{90}}{7.3s}\right)^{-1}\,.
\eary
%
%\subsubsection{SSC emission}
%
As one can see the synchrotron self-absorption is in the weak absorption regime ($E^{syn}_{\gamma, a,r} < E^{syn}_{\gamma, c,r}$), hence there is not a thermal component besides the power-law spectrum. The synchrotron emission from reverse shock predicts a peak of the flux density  at $\sim$ 100 eV.   On the other hand, accelerated electrons can upscatter photons from low to high energies as ${\small E^{\rm ssc}_{\rm \gamma,a,r}\sim2\gamma^2_{e,m,r}E^{syn}_{\rm \gamma, a,r}}$, ${\small E^{\rm ssc}_{\rm \gamma,m,r}\sim2\gamma^2_{e,m,r}E^{syn}_{\rm \gamma, m,r}}$, ${\small E^{ssc}_{\rm \gamma,c,r}\sim2\gamma^2_{\rm e,c,r}\,E^{syn}_{\gamma,c,r}}$ and ${\small F^{ssc}_{\rm \gamma,max,r}\sim\,k\tau\,F^{syn}_{\rm \gamma,max,r}}$  where {\small $k=4(p-1)/(p-2)$}, {\small $\tau=\frac{\sigma_T N(\gamma_e)}{4\pi r_d}$} is the optical depth of the shell and $N_e$ is the number of radiating electrons. From eq. (\ref{synrev_a}),  SSC  spectral breaks can be written as     
\bary\label{ssc_a}
E^{ssc}_{\rm \gamma,m,r}&\simeq& 103.55 \, {\rm \ MeV}\,  \biggl(\frac{1+z}{4}\biggr)^{-1}\,\epsilon_{e,r,-0.4}^4\,\epsilon_{B,r,-0.55}^{1/2}\,\Gamma^{4}_{2.72}\,  A^{3/2}_{\star,-1}\,E^{-1}_{54},\cr
E^{ssc}_{\rm \gamma,c,r}&\simeq&  5.86\times 10^{-3} \, {\rm \ eV}\, \biggl(\frac{1+z}{4}\biggr)^{-3/2}\,\biggl(\frac{1+x_r+x_r^2}{3}\biggr)^{-4}\,\epsilon_{B,r,-0.55}^{-7/2}\,\Gamma^{-6}_{2.72}\,A^{-6}_{\star,-1}\,E^{5/2}_{54}\,\left(\frac{T_{90}}{7.3s}\right)^{1/2},\cr
F^{ssc}_{\rm \gamma,max,r}&\simeq& 1.42\times10^2 \,{\rm \,Jy}\, \biggl(\frac{1+z}{4}\biggr)^{3}\,\epsilon_{B,r,-0.55}^{1/2}\,\Gamma^{-2}_{2.72}\,A^{3/2}_{\star,-1}\,D^{-2}_{28}\,E_{54}\,\left(\frac{T_{90}}{7.3s}\right)^{-2}\,.
\eary
From  SSC LCs in the fast-cooling regime of reverse shock \cite{2003ApJ...589L..69L},  eq. (\ref{ssc_a}) and the crossing time $\sim T_{90}/6$, one can describe the brightest LAT peak as SSC emission from reverse shock. In fact, the value of equipartition parameters $\epsilon_{B,r}=$0.28 and $\epsilon_{e,r}=$0.4 were obtained after fitting this peak at 5.5 s (see \cite{Fraija}).  %
\section{Conclusions}
We have presented a leptonic model based on the evolution of an early afterglow in the stellar wind. The current model having six free parameters (bulk Lorentz factor, density of the stellar wind, electron and magnetic equipartition parameters) accounts for the main temporal and spectral characteristics of GRB 110731A. The LAT emission extended up to 853 s is interpreted as synchrotron emission from forward shock  and the brightest LAT peak is described evoking the SSC emission from reverse shock. Additionally, X-ray and optical fluxes are naturally explained by synchrotron emission from forward shock.  
We consider that the ejecta propagating in the stellar wind is early decelerated at $\sim 5.5$ s and the reverse shock evolves in the thick-shell regime.   Taking into account  the values of redshift  $z=2.83$ \cite{2011GCN..12225...1T}, energy $E\simeq10^{54}$ erg, duration of GRB $T_{90}$= 7.3 s  \cite{2013ApJ...763...71A,2011GCNR..343....1O} and the stellar wind $A=5.0\times 10^{10}$ g/cm \cite{2000ApJ...536..195C}, we get that the value of the Lorentz factor is $\Gamma\simeq 520$.\\
Comparing the magnetic equipartition parameters that best describe the emission at forward and reverse shocks, we can see that magnetic fields in both shocks are related by  $B_f\simeq2\times 10^{-2}\,B_r$. The previous result as found in other bursts \cite{2003ApJ...597..455K, 2012ApJ...751...33F,2012ApJ...755..127S} illustrates that the magnetic field in the reverse-shock region is stronger ($\sim$ 50 times) than in the forward-shock region which indicates that the ejecta is magnetized. 
Although from this burst photons at sub-TeV energies were not observed, in other bursts they might be present. In this case, these sub-TeV could be modeled as SSC emission from forward shock and also be candidates to be detected by TeV $\gamma$-ray observatories as the High Altitude Water Cherenkov observatory (HAWC) \cite{2015ApJ...800...78A}.

\end{document}